\documentclass[11pt,twoside]{article}
\usepackage{asp2004}
\usepackage{psfig}
\usepackage{epsf}
\usepackage{graphics}
\usepackage{lscape}
\begin{document}
\title{The outflowing disks of B[e] supergiants and unclassified B[e] stars}
\author{M. Kraus}
\affil{Astronomical Institute, Utrecht University, Princetonplein 5, 3584 CC Utrecht, The Netherlands}
\author{M. Borges Fernandes}
\affil{Observat\'{o}rio Nacional-MCT, Rua General Jos\'{e} Cristino 77, 20921-400 S\~{a}o Cristov\~{a}o, Rio de Janeiro, Brasil}

\begin{abstract}
B[e] supergiants are known to possess outflowing cool disks but also some 
unclassified B[e] stars show clear indications for the presence of a 
neutral disk. We derive constraints on the disk mass loss rates, temperature 
distributions and disk opening angles for the Small Magellanic Cloud B[e] 
supergiant Hen S 18 and the unclassified galactic B[e] star Hen 2-90 by 
modeling the line luminosities of the [O{\sc i}] lines arising in their optical
spectrum. These lines are supposed to form in a hydrogen neutral disk. We find 
disk mass fluxes of order $3.4\times 10^{-4}$\,g\,s$^{-1}$cm$^{-2}$ and 
$5.5\times 10^{-1}$\,g\,s$^{-1}$cm$^{-2}$ resulting in disk mass loss rates of
$1.0\times 10^{-4}$\,M$_{\sun}$yr$^{-1}$ and $1.5\times 
10^{-5}$\,M$_{\sun}$yr$^{-1}$ for Hen S 18 and Hen 2-90, respectively.
\end{abstract}
\thispagestyle{plain}

\section{Introduction}

The group of stars showing the B[e] phenomenon is heterogeneous and has been 
divided by \citet{l1998} into subgroups according to their 
evolutionary phase. These subgroups contain supergiants, Herbig stars, 
symbiotic objects and compact planetary nebulae. The biggest group, however, 
are the unclassified B[e] stars whose evolutionary phase is not or not 
unambiguously known.

The optical spectra\footnote{Based on observations with the 1.52m telescope
at the European Southern Observatory (La Silla, Chile), under the agreement 
with the Observat\'{o}rio Nacional-MCT (Brasil)} of the Small Magellanic Cloud 
(SMC) B[e] supergiant Hen S 18 and of the galactic unclassified B[e] star 
Hen 2-90 show both the presence of very strong emission in the [O{\sc i}] lines 
which indicates that there must be a huge amount of neutral material close to 
the star.

In a recent paper, \citet{k2003} showed that the disks around B[e] supergiants 
can indeed become neutral, i.e. hydrogen can recombine, even close to the hot 
stellar surface, simply due to the high equatorial mass fluxes of these stars 
that result in effective shielding of the disk material from the ionizing 
stellar continuum photons.

\begin{figure}[!ht]
\plotone{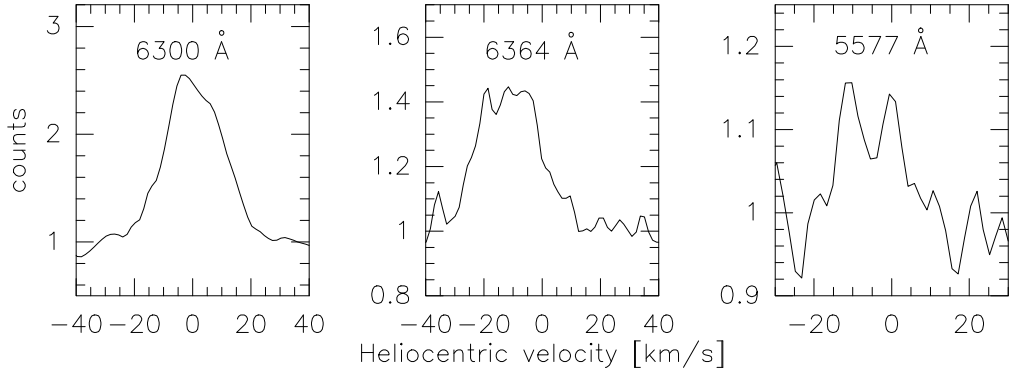}
\caption{Heliocentric velocities of the three [O{\sc i}] lines in the FEROS
spectrum of Hen S 18. The line wings of about 25\,km\,s$^{-1}$ indicate the
disk outflow velocity projected to the line of sight. The real outflow 
velocity is somewhat higher since Hen S 18 is seen under an intermediate
inclination angle.}
\label{velo_S18}
\end{figure}
                                                                                
\begin{figure}[!ht]
\plotone{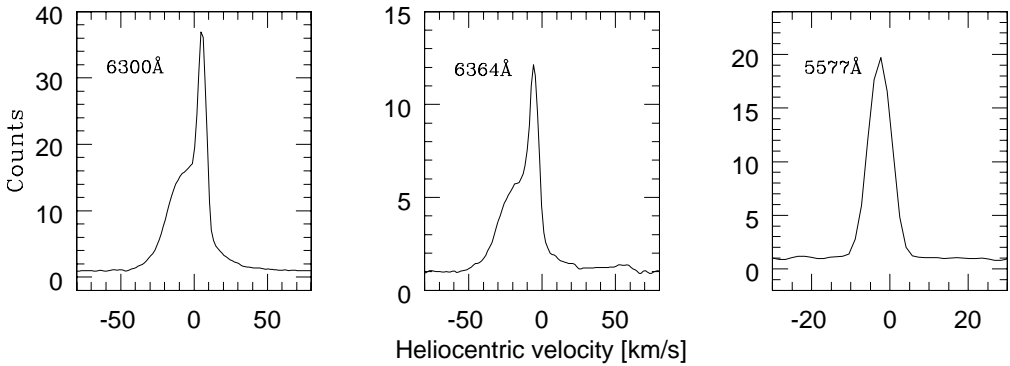}
\caption{Same as Fig.\,\ref{velo_S18}, but for Hen 2-90. This system is seen
edge-on. The wings of about 35\,km\,s$^{-1}$ indicate the outflow velocity.}
\label{velo_2-90}
\end{figure}

\section{The outflowing disk model}
                                                                                
Emission of O{\sc i} is expected to arise from regions in which
hydrogen is neutral due to the about equal ionization potentials of H and O.
The best location is therefore the outflowing disk.
To simplify the model calculations we assume that the outflowing disk is
neutral in hydrogen already at the stellar surface. The only free
electrons available to collisionally excite the levels in O{\sc i}
result from elements like Fe which have a much lower ionization potential
than H. The electron density is therefore of order $N_{\rm e}(r)\simeq
10^{-4}\ldots 5\times 10^{-4}\,N_{\rm H}(r)$ depending on metallicity and
on the internal ionization structure, i.e. temperature distribution, of the
disk. The radial hydrogen density distribution is given by the equation of
mass continuity. The terminal velocities for each star are derived from the
wings of their [O{\sc i}] lines (see Figs.\,\ref{velo_S18} and \ref{velo_2-90})
assuming that Hen S 18 is seen under an intermediate angle and Hen 2-90 is seen
edge-on.

We calculate the level population by solving
the statistical equilibrium equations in a 5-level atom.
Since the forbidden lines are optically thin, no radiation transfer needs
to be calculated which simplifies our analysis. There are three [O{\sc i}] 
lines in our spectra of which we model the luminosities. These lines have 
laboratory wavelengths of 5577\,\AA,~6300\,\AA,~and 
6364\,\AA~(see Figs.\,\ref{velo_S18} and \ref{velo_2-90}).

\section{The SMC B[e] supergiant Hen S 18}

\begin{figure}[!t]
\plotone{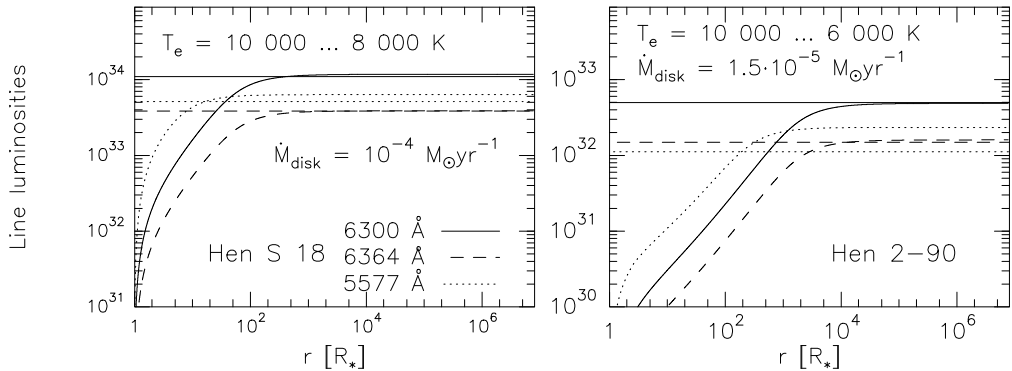}
\caption{Luminosities of the [O{\sc i}] lines of the SMC B[e] supergiant
Hen S 18 (left) and the unclassified galactic B[e] star Hen 2-90 (right). 
The straight lines are the observed values, the curved lines represent
the modeled luminosities as a function of radial distance from the star.
The identification of the lines is the same in both plots.}
\label{fits}
\end{figure}

Hen S 18 is a supergiant with effective temperature $T_{\rm eff}\simeq
25\,000$\,K, a luminosity of $\log (L/L_{\odot}) \simeq 5.3$ and a radius 
of about $R_{*} \simeq 39$\,R$_{\odot}$ \citep{l1998}. Its distance is 
roughly 60\,kpc. The oxygen abundance is set to $0.25\times {\rm solar}$, 
which is a mean SMC value. In Fig.\,\ref{fits} we show results for the line 
luminosity calculations for the [O{\sc i}] lines indicated. 
We need a disk mass flux of $3.4\times 10^{-4}$g\,s$^{-1}$cm$^{-2}$
which results into a disk mass loss rate of $\dot{M}_{\rm disk}
= 1.0\cdot 10^{-4}$M$_{\odot}$yr$^{-1}$ if we assume that the 
disk covers a fraction of about 0.2 of the total volume.

We want to strengthen that this is a lower limit to the disk (and 
therefore the total) mass loss rate of the star 
because we used the typical SMC abundance in our calculations. Since 
supergiants are normally in an evolved phase, the surface oxygen abundance
might be much lower due to several dredge-ups. An underabundance in O
would then result in a much higher mass flux needed to explain the 
observed line luminosities. 

\section{The unclassified B[e] star Hen 2-90}

Hen 2-90 has been classified either as a symbiotic object or as a compact 
planetary nebula. Its HST image \citep{s2002} reveals a bipolar high-ionized
wind, a low-ionized wind at intermediate latitudes as well as a high-density
circumstellar disk. In addition, a bipolar jet has been found with
several knots extending up to $\sim 10\arcsec$~on both sides of the star.
The clearly distinct regions of different ionization degrees leads us
to the assumption that Hen 2-90 has either a latitude dependent surface
temperature being hotter on the poles, or a latitude dependent mass flux
being strongest at the equator, or both.
                                                                
The star is at a distance of about 2\,kpc and the following stellar parameters
are known: $T_{\rm eff} \simeq 50\,000$\,K, $R_{*}\simeq 0.38$\,R$_{\odot}$
and $\log(L/L_{\odot})\simeq 3$ \citep{c1993}. In Fig.\,\ref{fits} we show
the modeled line luminosities. The disk mass flux is found to be of order 
$5.5\times 10^{-1}$g\,s$^{-1}$cm$^{-2}$. From the HST image we find that 
the disk covers about 0.2 of the wind volume leading to a disk mass loss 
rate of about $1.5\cdot 10^{-5}$M$_{\odot}$yr$^{-1}$. 

For this star we did not only model the [O{\sc i}] lines but almost 
all available forbidden lines arising in the optical spectrum 
\citep{k2004}. These lines come from all the different ionization regions
in the non-spherical wind seen in the HST image. From a self-consistent
modeling we find that the star must be underabundant in C, N, and also
in O with an O abundance of only $0.3\times {\rm solar}$. We could
explain the different ionization regions indeed in terms of a latitude
dependent mass flux as well as a latitude dependent surface temperature
which might be explained in terms of a rapidly rotating underlying star.
In addition, we could fix the total mass loss rate of Hen 2-90 to about 
$3\times 10^{-5}$\,M$_{\sun}$yr$^{-1}$.

\section{Discussion and Conclusions}

It is obvious that our model predicts for both stars a [O{\sc i}] 
5577\,\AA~luminosity which is higher than the observed value.
This line corresponds to the transition $5\longrightarrow 4$ in our adopted
5-level atom. There exists one single permitted transition between its upper level
and an energetically much higher lying level with wavelength $\lambda = 
1217.6$\,\AA~which falls into the wavelength range covered by a broadened 
Ly\,$\alpha$ line ($\lambda_{\rm Ly\,\alpha} = 1215.6$\,\AA). The fifth level might
therefore be depopulated radiatively into this higher state from which several
permitted lines arise. Consequently, the observable 5577\,\AA~line luminosity
will decrease. This depopulation mechanism might also explain why the 5577\,\AA~line
is much narrower than the other two [O{\sc i}] lines in our sample.

Nevertheless, the presence of [O{\sc i}] lines proofs the existence of cool 
and neutral material close to hot B[e] stars. From modeling the line 
luminosities we could (i) fix a temperature distribution within the disk
and (ii) determine the disk mass fluxes resulting in disk mass loss rates
which are lower limits to the total mass loss rates for the two studied stars.

\begin{acknowledgements}
M.K. acknowledges financial support from the Nederlandse Organisatie voor
Wetenschappelijk Onderzoek grant No.\,614.000.310.
M.B.F. acknowledges financial support from CAPES (Ph.D. studentship).
\end{acknowledgements}


\begin{thebibliography}{}
\bibitem[Costa et al.(1993)]{c1993} Costa, R. D. D., de Freitas Pacheco,
  J. A. \& Maciel, W. J. 1993, A\&A, 276, 184
\bibitem[Kraus et al.(2004)]{k2004} Kraus, M., Borges Fernandes, M., 
  de Ara\'{u}jo, F. X. \& Lamers, 
  H. J. G. L. M. 2004, A\&A, {\it submitted}
\bibitem[Kraus \& Lamers(2003)]{k2003} Kraus, M. \& Lamers, H. J. G. L. M., 
  2003, A\&A, 405, 165
\bibitem[Lamers et al.(1998)]{l1998} Lamers, H. J. G. L. M., Zickgraf, F.-J.,
  de Winter, D., Houziaux, L. \& Zorec, J. 1998, A\&A, 340, 117
\bibitem[Sahai et al.(2002)]{s2002} Sahai, R., Brillant, S., Livio, M., Grebel, 
  E. K., Brandner, W., Tingay, S. \& Nyman, L.-\AA. 2002, ApJ, 573, L\,123
\end{thebibliography}
\end{document}